\documentclass[preprint,superscriptaddress,nofootinbib]{revtex4}
\usepackage[dvips]{graphicx}
\usepackage{amsmath,amssymb,color}

\newcommand{\MNS}{{\text{MNS}}}

\newcommand{\eV}{{\text{eV}}}

\newcommand{\GeV}{{\text{GeV}}}
\newcommand{\TeV}{{\text{TeV}}}

\newcommand{\BR}{\text{BR}}

\newcommand{\U}{{\text{U}}}
\newcommand{\SU}{{\text{SU}}}

\newcommand{\wh}[1]{\widehat{#1}}
\newcommand{\wt}[1]{\widetilde{#1}}
\newcommand{\ellp}{\ell^\prime}

\allowdisplaybreaks

\begin{document}

\preprint{UT-HET 104}
\preprint{KU-PH-018}

\title{
R-Parity Conserving Supersymmetric Extension of the Zee Model
}

\author{Shinya Kanemura}
\email{kanemu@sci.u-toyama.ac.jp}
\affiliation{
Department of Physics,
University of Toyama,
3190 Gofuku,
Toyama 930-8555, Japan
}
\author{Tetsuo Shindou}
\email{shindou@cc.kogakuin.ac.jp}
\affiliation{
Division of Liberal-Arts,
Kogakuin University,
1-24-2 Nishi-Shinjuku,
Tokyo 163-8677, Japan
}
\author{Hiroaki Sugiyama}
\email{sugiyama@sci.u-toyama.ac.jp}
\affiliation{
Department of Physics,
University of Toyama,
3190 Gofuku,
Toyama 930-8555, Japan
}


\begin{abstract}
 We extend the Zee model,
where tiny neutrino masses are
generated at the one-loop level,
to a supersymmetric model
with R-parity conservation.
 It is found that
the neutrino mass matrix
can be consistent with the neutrino oscillation data
thanks to the nonholomorphic Yukawa interaction
generated via one-loop diagrams of sleptons.
 We find a parameter set of the model,
where in addition to the neutrino oscillation data,
experimental constraints
from the lepton flavor violating decays
of charged leptons
and current LHC data
are also satisfied.
 In the parameter set,
an additional CP-even neutral Higgs boson
other than the standard-model-like one,
a CP-odd neutral Higgs boson,
and two charged scalar bosons
are light enough to be produced 
at the LHC and future lepton colliders.
 If the lightest charged scalar bosons
are mainly composed of the $\SU(2)_L$-singlet scalar boson in the model,
they would decay into $e\nu$ and $\mu\nu$
with 50\,\% of a branching ratio for each.
 In such a case,
the relation among the masses of
the charged scalar bosons and the CP-odd Higgs
in the minimal supersymmetric standard model
approximately holds with a radiative correction.
 Our model can be tested
by measuring the specific decay patterns of
charged scalar bosons
and the discriminative mass spectrum of additional scalar bosons.
\end{abstract}

\maketitle

\section{Introduction}

 Neutrino oscillation data%
~\cite{solar,Aharmim:2011vm,Gando:2010aa,Wendell:2010md,
acc-disapp,Abe:2014ugx,s-reac,An:2015rpe,Abe:2013hdq}
have indicated the existence of tiny masses of neutrinos,
which are absent in the standard model~(SM)
of particle physics.
 If the tiny neutrino masses
are generated by a new physics at a very high energy scale%
~(e.g., the seesaw mechanism~\cite{Ref:seesaw}),
such a new physics~(heavy new particles)
is not directly accessible by experiments.
 In contrast,
in scenarios based on
radiative generation of tiny neutrino masses%
~\cite{Zee:1980ai, ZB,
 Krauss:2002px, Ma:2006km, Aoki:2008av, Gustafsson:2012vj},
the smallness of neutrino masses
is deduced by the quantum effect
without introducing very heavy new particles.
 Therefore,
such scenarios can consequently be tested
by current and future collider experiments.

 The model proposed by A.~Zee~\cite{Zee:1980ai}
is the first model of radiative generation of neutrino masses,
where an $\SU(2)_L$-singlet scalar field with a hypercharge $Y=1$
and the second $\SU(2)_L$-doublet Higgs field with $Y=1/2$
are introduced to construct the one-loop diagram for the neutrino mass.
 Studies on the phenomenology in the Zee model
can be found in Refs.~\cite{Wolfenstein:1980sy,Zee,
Smirnov:1996bv, Cheung:1999az, Kanemura:2000bq, Balaji:2001ex,
Zee_twozeros, He:2003ih, He:2011hs}.
 In the minimal version of the model (the so-called Zee-Wolfenstein 
model~\cite{Wolfenstein:1980sy}),
lepton flavor violating~(LFV) Yukawa couplings
with the second Higgs doublet are forbidden
at the tree level.
 However,
such a model has already been excluded
by current neutrino oscillation data%
~(See e.g., Ref.~\cite{He:2003ih}).
In order to reproduce the neutrino data, 
lepton flavor violating interactions are necessary%
~\cite{Balaji:2001ex, He:2003ih, He:2011hs}. 

 In the Zee model,
there are several problems to be solved.
 Although the Zee model with LFV couplings
is phenomenologically acceptable, 
they should be well controlled by some principle
in order to suppress the dangerous flavor 
changing neutral current~(FCNC) processes%
\footnote
{
 It would be also required to introduce
the third $\SU(2)_L$-doublet scalar field
which is only for the quark Yukawa interactions
without FCNC in the quark sector.
}\,%
\footnote{
 An idea to control such FCNC well so
that stringent constraints on
$\mu \to \overline{e}ee$ and $\mu \to e\gamma$
are automatically satisfied
is introducing the $A_4$ symmetry to the 
Zee model~\cite{Fukuyama:2010ff}.
}.
 The model is also confronted by the quadratic divergence problem 
like the SM\@. 
 In addition,
there is no dark matter~(DM) candidate
in the Zee model.
 If we consider a supersymmetric~(SUSY) extension of the Zee model,
we may be able to solve these problems simultaneously. 
 The quadratic divergence is automatically 
cancelled by the loop contribution of SUSY partner particles.
 If the R-parity is conserved,
the lightest SUSY particle becomes stable
and it can be DM\@.
 Moreover,
the LFV Yukawa couplings can naturally be induced.

 The previous study of the SUSY extension of the Zee model
is found in Ref.~\cite{Cheung:1999az},
where the R-parity  violation is introduced to
the Minimal SUSY SM~(MSSM).
 In this model,
right-handed sleptons play the role of the charged singlet scalar in 
the Zee model.
Since the sleptons carry lepton flavors in contrast with the singlet 
scalar in the Zee model, the flavor structure of the generated neutrino 
mass matrix becomes different from the one in the Zee-Wolfenstein model.
 SUSY models for the other scenarios of radiative neutrino masses
can be found in 
Refs.~\cite{Ma:2006uv, Aoki:2010ib, Kanemura:2013uva, Kanemura:2014cka}.

 In this paper,
we propose a SUSY extension of the Zee model
with the R-parity conservation~(the SUSY Zee model).
 The stability of the dark matter candidate is guaranteed.
 This is the simplest alternative example of the MSSM
with the right-handed neutrino superfields~(the SUSY seesaw model).
 In the SUSY Zee model,
the MSSM is extended by introducing a pair of 
$\SU(2)_L$-singlet superfields with hypercharge $Y=1$ and $-1$,
which also carry lepton numbers.
 The Higgs sector of the MSSM is
the type~II two Higgs doublet model~\cite{Gunion:1989we}
at the tree level.
 The extra Higgs doublet can 
play the role of the second Higgs doublet of the Zee model.
 In SUSY models,
nonholomorphic Yukawa interactions are generally induced
by the one-loop effect of SUSY particles%
~\cite{ref:Babu-Kolda, Brignole:2004ah, Kanemura:2004cn,
Paradisi:2006jp, Endo:2015oia, Kakizaki:2015zva}.
 This mechanism may be utilized
for generating the flavor violating interaction,
which is required for the Zee model to satisfy the neutrino data.
 The structure of the LFV Yukawa matrix is determined
by the flavor structure of the soft SUSY breaking slepton mass matrices. 
 Such radiatively induced coupling constants
are expected to be much smaller than
the other Yukawa coupling constants.
 In this model,
we study the neutrino mass matrix,
and we find a benchmark point for model parameters,
which satisfies the required structure of the neutrino mass matrix
and the constraints from experimental searches
for the LFV decays of charged leptons.
 On the benchmark point,
phenomenological consequences of our model are discussed,
and testability of our model at the LHC
and future lepton colliders such as
the International Linear Collider~(ILC)
is mentioned.

 This paper is organized as follows.
 In Sec.~\ref{sec:model}, our model is defined.
 Section~\ref{sec:mnu} is devoted
to showing how the neutrino masses are generated
at the one-loop level.
 In the section,
a benchmark set of model parameters
which satisfy neutrino oscillation data
are given.
 Phenomenology in the SUSY Zee model with the benchmark set
is discussed in Sec.~\ref{sec:pheno}.
 Conclusions are given in Sec.~\ref{sec:concl}.

\section{The Model}
\label{sec:model}

\begin{table}
\begin{tabular}{c||c|c||c|c||c}
 {}
  & Spin $0$
  & Spin $1/2$
  & $\SU(2)_L$
  & $\U(1)_Y$
  & Lepton \#
\\\hline\hline
 $\widehat{L}_\ell$
  & $
     \widetilde{L}_\ell
     =
     \begin{pmatrix}
      \widetilde{\nu}_{\ell L}^{}\\
      \widetilde{\ell}_L
     \end{pmatrix}
    $
  & $
     L_\ell
     =
     \begin{pmatrix}
      \nu_{\ell L}^{}\\
      \ell_L
     \end{pmatrix}
    $
  & ${\bf 2}$
  & $\displaystyle -\frac{1}{\,2\,}$
  & $1$
\\\hline
 $\widehat{\ell}^c$
  & $\widetilde{\ell}_R^\ast$
  & $(\ell_R)^c$
  & ${\bf 1}$
  & $1$
  & $-1$
\\\hline
 $\widehat{\Phi}_u$
  & $
     \Phi_u
     =
     \begin{pmatrix}
      \phi_u^+\\
      \phi_u^0
     \end{pmatrix}
    $
  & $
     \widetilde{\Phi}_u
     =
     \begin{pmatrix}
      \widetilde{\phi}_u^+\\
      \widetilde{\phi}_u^0
     \end{pmatrix}
    $
  & ${\bf 2}$
  & $\displaystyle \frac{1}{\,2\,}$
  & $0$
\\\hline
 $\widehat{\Phi}_d$
  & $
     \Phi_d
     =
     \begin{pmatrix}
      \phi_d^0\\
      \phi_d^-
     \end{pmatrix}
    $
  & $
     \widetilde{\Phi}_d
     =
     \begin{pmatrix}
      \widetilde{\phi}_d^0\\
      \widetilde{\phi}_d^-
     \end{pmatrix}
    $
  & ${\bf 2}$
  & $\displaystyle -\frac{1}{\,2\,}$
  & $0$
\\\hline
 $\widehat{\omega}_1^+$
  & $\omega_1^+$
  & $\widetilde{\omega}_1^+$
  & ${\bf 1}$
  & $\displaystyle 1$
  & $-2$
\\\hline
 $\widehat{\omega}_2^-$
  & $\omega_2^-$
  & $\widetilde{\omega}_2^-$
  & ${\bf 1}$
  & $\displaystyle -1$
  & $2$
\end{tabular}
\caption
{
 Superfields of the SUSY Zee model.
 The baryon number is zero for all particles in this table.
}
\label{tab:particle}
\end{table}

 Superfields of the SUSY Zee model
are partially listed in Table~\ref{tab:particle}.
 The transformation property under the R-parity
is given by $(-1)^{3(B-L)+2s}$,
where $B$~($L$) is the baryon~(lepton) number
and $s$ denotes the spin.
 Component fields with "tilde" are odd under the R-parity.
 The relevant part of the superpotential is constructed as
\begin{align}
 \mathcal{W}=&
 y_\ell^{}\,
 \wh{\ell}^c\, \wh{L}_\ell^T (-i\sigma_2) \wh{\Phi}_d
 + (Y_A^{(0)})_{\ell\ellp}\,
   \wh{L}_\ell^T (i\sigma_2) \wh{L}_{\ellp}\, \wh{\omega}_1^+
 + \mu_\Phi^{}\, \wh{\Phi}_d^T (i\sigma_2) \wh{\Phi}_u
 + \mu_{\omega}^{}\, \wh{\omega}_1^+\, \wh{\omega}_2^-\;,
\label{eq:sup-pot}
\end{align}
where $(Y_A^{(0)})^T = -Y_A^{(0)}$,
and $\sigma_i$~$(i = 1\text{-}3)$ are the Pauli matrices.
 The relevant part of the soft-SUSY breaking terms is given by 
\begin{align}
	\mathcal{L}_{\text{soft}}=&
	- m_{\Phi_u}^2 \Phi_u^{\dagger} \Phi_u
	- m_{\Phi_d}^2 \Phi_d^{\dagger} \Phi_d
	- \left\{
           B_\Phi\, \mu_\Phi^{} \Phi_d^T (i\sigma_2) \Phi_u + \text{h.c.}
          \right\}
	\nonumber\\
	&
        - m_{\omega_1}^2 \omega_1^+ \omega_1^-
	- m_{\omega_2}^2 \omega_2^+ \omega_2^-
	- \left\{
           B_\omega\, \mu_\omega^{}\, \omega_1^+ \omega_2^-
           + \text{h.c.}
          \right\}
	\nonumber\\
	&
	- (m_{\wt{L}}^2)_{\ell\ellp} \wt{L}_\ell^{\dagger} \wt{L}_{\ellp}
	- (m_{\wt{\ell}}^2)_{\ell\ellp}\, \wt{\ell}^* \tilde{\ell}^\prime
	\nonumber\\
	&
	- \left\{
           y_\ell (A_E)_{\ell\ellp}\,
           \wt{\ell}^* \wt{L}_{\ellp}^T (i\sigma_2) \Phi_d
           + \text{h.c.}
          \right\}
	- \left\{
           (A_\omega)_{\ell\ellp}\,
           \wt{L}_\ell (i\sigma_2) \wt{L}_{\ellp}\, \omega_1^+
           + \text{h.c.}
          \right\}
	\nonumber\\
	&
	-\left\{
          C_1 \Phi_u^{\dagger} \Phi_d\, \omega_1^+
          + C_2 \Phi_d^{\dagger} \Phi_u\, \omega_2^-
          + \text{h.c.}
         \right\}\;,
\end{align}
where $A_\omega^T = -A_\omega$.
 Notice that the term of $C_1$
gives the important interaction required in the non-SUSY Zee model
as the source of the lepton number violation by two units.%
\footnote{
 Such three-scalar terms
with both fields and conjugate fields
are the so-called C-terms~\cite{Hall:1990ac},
which are usually ignored
because most SUSY breaking scenarios
do not generate the C-terms.
 However,
the C-terms can be generated
in some SUSY breaking scenarios,
e.g., in an intersecting D-brane model
with the flux compactification~\cite{Camara:2003ku}.
 Some detailed discussion about
the SUSY breaking effects for the radiative neutrino mass
can be found, e.g., in Ref.~\cite{Figueiredo:2014gpa}.
}

 In order to generate the neutrino mass matrix
at the one-loop level,
the following Yukawa interactions are used:
\begin{eqnarray}
&&
{\mathcal L}_{\text{Yukawa}}
 =
 y_\ell^{}\,
 \overline{\ell_R}\, L_{\ellp}^T (-i\sigma_2) \Phi_d
 +
 (Y_2)_{\ell\ellp}\,
 \overline{\ell_R}\, L_{\ellp}^T \Phi_u^\ast
\nonumber\\
&&\hspace*{20mm} {}
 +
 (Y_A^{(0)})_{\ell\ellp}\,
 L_\ell^T (i\sigma_2) L_{\ellp}\, \omega_1^+
 +
 (Y_{2A})_{\ell\ellp}\,
 L_\ell^T (i\sigma_2) L_{\ellp}\, \omega_2^+ ,
\label{eq:Yukawa-1}
\end{eqnarray}
where the first and the third terms
are obtained from the superpotential $\mathcal{W}$
in eq.~\eqref{eq:sup-pot}.
 The second and the fourth terms
are generated at the one-loop level%
~\cite{ref:Babu-Kolda, Brignole:2004ah, Paradisi:2006jp, Kanemura:2004cn,
Endo:2015oia, Kakizaki:2015zva}.
 In eq.~\eqref{eq:Yukawa-1},
$Y_{A2}$ can be ignored,
when $\omega_2^+$ is very heavy.
 We here consider such a case.
 The Yukawa matrix $Y_2$ is generated
through the slepton mixing as follows%
~\cite{ref:Babu-Kolda, Brignole:2004ah, Paradisi:2006jp, Kanemura:2004cn,
Endo:2015oia, Kakizaki:2015zva}:
\begin{eqnarray}
 (Y_2)_{\ell\ellp}
 &=&
  y_\ell^{}
  \Bigl\{
   (\epsilon_1)_\ell \delta_{\ell\ellp}^{}
   + (\epsilon_2)_{\ell\ellp}
  \Bigr\} ,
\\
%
%
 (\epsilon_1)_\ell
 &=&
  - \frac{\alpha_1}{8\pi} \mu_\Phi^{} M_1
    \biggl[
     2 I_3( M_1^2 , m_{\wt{\ell}_L^{}}^2 , m_{\wt{\ell}_R^{}}^2 )
     + I_3( M_1^2 , \mu_\Phi^2 , m_{\wt{\ell}_L^{}}^2 )
     - 2 I_3( M_1^2 , \mu_\Phi^2 , m_{\wt{\ell}_R^{}}^2 )
    \biggr]
\nonumber\\
 &&{}
  + \frac{\alpha_2}{8\pi} \mu_\Phi^{} M_2
    \biggl[
     I_3( M_2^2 , \mu_\Phi^2 , m_{\wt{\ell}_L^{}}^2 )
     + 2 I_3( M_2^2 , \mu_\Phi^2 , m_{\wt{\nu}_{\ell L}^{}}^2 )
    \biggr] ,
\label{eq:ep1}\\
%
%
 (\epsilon_2)_{\ell\ellp}
 &=&
  - \frac{\alpha_1}{8\pi} \mu_\Phi^{} M_1
    \bigl( \Delta m_{\wt{L}}^2 \bigr)_{\ell\ellp}
    \biggl[
     2 I_4( M_1^2 , m_{\wt{\ell}_L^{}}^2 ,
            m_{\wt{\ell}_R^{}}^2 , m_{\wt{\ell}_L^\prime}^2 )
     + I_4( M_1^2 , \mu_\Phi^2 ,
            m_{\wt{\ell}_L^{}}^2 , m_{\wt{\ell}_L^\prime}^2 )
    \biggr]
\nonumber\\
 &&{}
  - \frac{\alpha_1}{8\pi} \mu_\Phi^{} M_1
    \bigl( \Delta m_{\wt{\ell}}^2 \bigr)_{\ell\ellp}
    \biggl[
     2 I_4( M_1^2 , m_{\wt{\ell}_L^{}}^2 ,
            m_{\wt{\ell}_R^{}}^2 , m_{\wt{\ell}_R^\prime}^2 )
     +
     2 I_4( M_1^2 , \mu_\Phi^2 ,
            m_{\wt{\ell}_R^{}}^2 , m_{\wt{\ell}_R^\prime}^2 )
    \biggr]
\nonumber\\
 &&{}
  + \frac{\alpha_2}{8\pi} \mu_\Phi^{} M_2
    \bigl( \Delta m_{\wt{L}}^2 \bigr)_{\ell\ellp}
    \biggl[
     I_4( M_2^2 , \mu_\Phi^2 ,
            m_{\wt{\ell}_L^{}}^2 , m_{\wt{\ell}_L^\prime}^2 )
     +
     2 I_4( M_2^2 , \mu_\Phi^2 ,
            m_{\wt{\nu}_{\ell L}^{}}^2 , m_{\wt{\nu}_{\ellp L}^{}}^2 )
    \biggr] ,
\label{eq:ep2}
\end{eqnarray}
where $M_1$~($M_2$) is the soft SUSY breaking mass
of $\U(1)_Y$~($\SU(2)_L$) gauginos.
 The matrices $\Delta m_{\wt{L}}^2$ and $\Delta m_{\wt{\ell}}^2$
denote off-diagonal parts of
$(m_{\wt{L}}^2)_{\ell\ellp}$ and $(m_{\wt{\ell}}^2)_{\ell\ellp}$,
respectively.
 Thus, $(\epsilon_2)_{\ell\ell} = 0$.
 Loop functions $I_3(x,y,z)$ and $I_4(x,y,z,w)$ are
defined as
\begin{eqnarray}
 I_3(x, y, z)
 &\equiv&
  - \frac{
          x y \ln(x/y) + y z \ln(y/z) + z x \ln(z/x)
         }
         {
          ( x - y )( y - z )( z - x )
         } ,
\\
%
%
 I_4(x, y, z, w)
 &\equiv&
  - \frac{ x \ln x }{ ( y - x )( z - x )( w - x ) }
  - \frac{ y \ln y }{ ( x - y )( z - y )( w - y ) }
\nonumber\\
&&{}\hspace*{5mm}
  - \frac{ z \ln z }{ ( x - z )( y - z )( w - z ) }
  - \frac{ w \ln w }{ ( x - w )( y - w )( z - w ) } .
\end{eqnarray}
 Although all terms in eq.~\eqref{eq:ep2}~(in eq.~\eqref{eq:ep1})
are proportional to $\mu_\Phi^{}$,
the first and the third terms in eq.~\eqref{eq:ep2}
~(the first term in eq.~\eqref{eq:ep1})
do not contain Higgsinos in the loop.
 Therefore,
sizable $\epsilon_2$~(and $\epsilon_1$ also)
can be generated by these terms
if we take much larger $\mu_\Phi$
than the other mass scales~(e.g., $M_1$, $m_{\wt{L}}^{}$).
 Then,
$\epsilon_1$ and $\epsilon_2$ are almost independent of the value of $M_2$.
 Yukawa interactions in eq.~\eqref{eq:Yukawa-1}
can be rewritten as
\begin{eqnarray}
{\mathcal L}_{\text{Yukawa}}
 =
 \frac{ \sqrt{2}\, m_\ell }{ v }\,
 \overline{\ell_R}\, L_\ell^T \Phi_v^\ast
 +
 \frac{ \sqrt{2}\, m_\ell }{ v }\,
 X_{\ell\ellp}\, \overline{\ell_R}\, L_{\ellp}^T \Phi_0^\ast
 +
 (Y_A)_{\ell\ellp}\,
 L_\ell^T (i\sigma_2) L_{\ellp}\, \omega_1^+ ,
\label{eq:Yukawa-2}
\end{eqnarray}
where $v^2 \equiv v_u^2 + v_d^2 = (246\,\GeV)^2$,
$v_u \equiv \sqrt{2} \langle \phi_u^0 \rangle$,
and $v_d \equiv \sqrt{2} \langle \phi_d^0 \rangle$.
 The matrix $Y_A$
is an arbitrary antisymmetric matrix.
 Two Higgs doublet fields $\Phi_0$ and $\Phi_v$
are defined as
\begin{eqnarray}
 \begin{pmatrix}
  \Phi_0\\
  \Phi_v
 \end{pmatrix}
 \equiv
  \begin{pmatrix}
   c_\beta & -s_\beta\\
   s_\beta &  c_\beta
  \end{pmatrix}
  \begin{pmatrix}
   \Phi_u\\
   (-i\sigma_2)\Phi_d^\ast
  \end{pmatrix} ,
\end{eqnarray}
where $s_\beta \equiv \sin\beta$ and
$c_\beta \equiv \cos\beta$
for $\tan\beta = v_u/v_d$.
 Interactions in eq.~\eqref{eq:Yukawa-2}
are expressed in terms of mass eigenstates of charged leptons%
\footnote{
 Charged leptons in eq.~\eqref{eq:Yukawa-1}
are not mass eigenstates
although we used the same notation $\ell$.
},
which are given by
diagonalizing the mass matrix
$c_\beta y_\ell^{} + s_\beta Y_2$.
 By keeping $\epsilon_1$ and $\epsilon_2$
up to linear terms
(keeping $\epsilon_1 \tan\beta$ for all order),
the matrix $X$ is given by~\cite{ref:Babu-Kolda}
\begin{eqnarray}
 X_{\ell\ellp}
 =
  - \tan\beta\, \delta_{\ell\ellp}
  + \frac{ 1 + \tan^2\beta }{( 1 + \tan\beta (\epsilon_1)_\ell )^2 }
    \Bigl\{
     (\epsilon_1)_\ell \delta_{\ell\ellp}
     + (\epsilon_2)_{\ell\ellp}
    \Bigr\} .
\end{eqnarray}
 The off-diagonal elements of $X$ provide FCNC,
and they are important to obtain
the appropriate structure of the neutrino mass matrix.

Since Nambu-Goldstone modes
are contained in $\Phi_v$,
mass eigenstates of the three charged bosons
are given by linear combinations of
$\omega_1^+$, $\omega_2^+$,
and a charged component $\phi_0^+$ of $\Phi_0$.
 The matrix of squared masses of charged bosons
is given in a basis of $(\phi_0^+, \omega_1^+, \omega_2^+)$ by
\begin{eqnarray}
M^2_{H^+}
&=&
 \begin{pmatrix}
  m_W^2 + m_A^2
   & \frac{ C_1 v }{\sqrt{2}}
   & \frac{ C_2 v }{\sqrt{2}} \\
  \frac{ C_1 v }{\sqrt{2}}
   & (M^2_{H^+})_{22}
   & B_\omega \mu_\omega^{} \\
 \frac{ C_2 v }{\sqrt{2}}
  & B_\omega \mu_\omega^{}
  & (M^2_{H^+})_{33}
 \end{pmatrix} ,
\\
%
%
(M^2_{H^+})_{22}
&=&
 - m_W^2 \tan^2\theta_W \cos(2\beta) + m_{\omega 1}^2 + \mu_\omega^2 ,
\\
%
%
(M^2_{H^+})_{33}
&=&
 m_W^2 \tan^2\theta_W \cos(2\beta) + m_{\omega 2}^2 + \mu_\omega^2 ,
\end{eqnarray}
where $m_A$ is the mass of the CP-odd Higgs boson $A^0$.
 The matrix $M^2_{H^+}$ is diagonalized as
$M^2_{H^+}
= U_{H^+} \text{diag}(m_{H_1^+}^2 , m_{H_2^+}^2 , m_{H_3^+}^2) U_{H^+}^\dagger$
with a unitary matrix $U_{H^+}$.
 We here assume that $(M^2_{H^+})_{33}$ is
much larger than the other elements for simplicity,
so that mixing effects via $C_2$ and $B_\omega$ can be ignored.
 The mass eigenvalues of charged scalar bosons are given by
\begin{eqnarray}
 m_{H_1^+}^2
 &=&
  \frac{1}{\,2\,}
  \left\{
   (M^2_{H^+})_{22} + m_W^2 + m_A^2
   - \sqrt{
      \Bigl( (M^2_{H^+})_{22} - m_W^2 - m_A^2 \Bigr)^2
      + 2 C_1^2 v^2
     }
  \right\} ,
\\
%
%
 m_{H_2^+}^2
 &=&
  \frac{1}{\,2\,}
  \left\{
   (M^2_{H^+})_{22} + m_W^2 + m_A^2
   + \sqrt{
      \Bigl( (M^2_{H^+})_{22} - m_W^2 - m_A^2 \Bigr)^2
      + 2 C_1^2 v^2
     }
  \right\} ,
\\
%
%
 m_{H_3^+}^2
 &=&
  (M^2_{H^+})_{33} .
\end{eqnarray}
 The mixing matrix is given by
\begin{eqnarray}
 U_{H^+}
  &=&
   \begin{pmatrix}
    \cos\theta_+ & -\sin\theta_+ & 0\\
    \sin\theta_+ & \cos\theta_+ & 0\\
    0 & 0 & 1
   \end{pmatrix} ,\\
%
%
 \sin^2{2\theta_+}
  &=&
   \frac{ 2 C_1^2 v^2 }
        { \left( m_{H_2^+}^2 - m_{H_1^+}^2 \right)^2 }
  =
   \frac{
          4
          ( m_{H_2^+}^2 - m_W^2 - m_A^2 )
          ( m_W^2 + m_A^2 - m_{H_1^+}^2 )
        }
        { \left( m_{H_2^+}^2 - m_{H_1^+}^2 \right)^2 } .
\end{eqnarray}
 For $\theta_+ \simeq 0$,
the charged scalar boson $H_1^+$
is the singlet-like one~($H_1^+ \simeq \omega_1^+$)
while $H_2^+$ is almost the same
as the charged Higgs boson of the MSSM\@.

\section{Neutrino Mass and Benchmark Scenario}
\label{sec:mnu}

\begin{figure}[t]
\begin{center}
\includegraphics[scale=0.7]{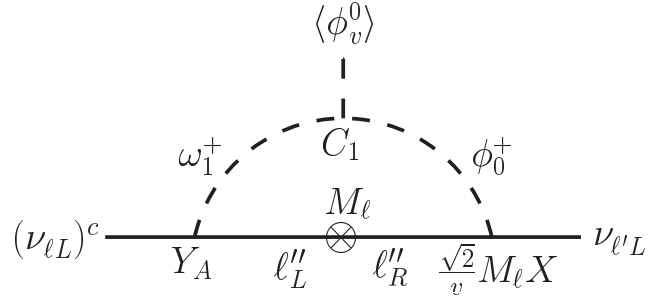}
\vspace*{-5mm}
\caption{
 The one-loop diagram for
light Majorana neutrino masses
in the SUSY Zee model.
}
\label{fig:mnu}
\end{center}
\end{figure}

 The neutrino mass matrix $(m_\nu^{})_{\ell\ellp}$
in the flavor basis,
$(1/2) (m_\nu^{})_{\ell\ellp} \overline{(\nu_{\ell L}^{})^c} \nu_{\ellp L}^{}
+ \text{h.c.}$,
is generated by the diagram in Fig.~\ref{fig:mnu}
and its transpose diagram.
 Keeping the leading terms of the charged lepton masses,
the neutrino mass matrix is expressed as
\begin{eqnarray}
m_\nu^{}
&=&
 \frac{\sqrt{2}}{v}
 C_\text{loop}
 \Bigl[
  Y_A M_\ell^2 X
  +
  (Y_A M_\ell^2 X)^T
 \Bigr] ,
\label{eq:mnu}
\end{eqnarray}
where $M_\ell \equiv \text{diag}(m_e, m_\mu, m_\tau)$
and $C_\text{loop}$ is
a flavor-independent factor
calculated by the loop integration.
 The explicit form of $C_\text{loop}$ is given by
\begin{eqnarray}
C_\text{loop}
=
 \frac{ \sin{2\theta_+} }{32\pi^2}\,
 \ln\frac{ m_{H_1^+}^2 }{ m_{H_2^+}^2 } .
\end{eqnarray}
 The case with $X=1$ corresponds to
the Zee-Wolfenstein model.
 Since we assume rather heavy SUSY particles
in order to satisfy constraints
from LFV decays of charged leptons,
we can ignore a contribution from
the $\wt{\omega}_1^+$-$\wt{\phi}_0^+$-$\wt{\ell}$ loop.%
\footnote{
 The interaction $\Phi_0^\dagger \Phi_v\, \omega_1^+$
in Fig.~\ref{fig:mnu}
is replaced with
a nonholomorphic Yukawa interaction
$\wt{\Phi}_0^\dagger \Phi_v\, \wt{\omega}_1^+$,
which is generated via a one-loop diagram
with $\Phi_0$, $\omega_1^+$, and the bino.
}

 A mass matrix $(m_\nu^{})_{\ell\ell^\prime}$
for Majorana neutrinos in the flavor basis
can be diagonalized by
the Maki-Nakagawa-Sakata~(MNS) matrix~\cite{Maki:1962mu}
$U_\MNS$ as
\begin{eqnarray}
m_\nu^{}
= U_\MNS^\ast\,
  \text{diag}(
    m_1 e^{i\alpha_{12}^{}} ,\,
    m_2 ,\,
    m_3 e^{i\alpha_{32}^{}}
  )\,
  U_\MNS^\dagger,
\end{eqnarray}
where $\alpha_{12}^{}$ and $\alpha_{32}^{}$ are
Majorana phases~\cite{Mphase}.
 The MNS matrix can be parametrized as
\begin{eqnarray}
U_\MNS
=
 \begin{pmatrix}
  1 & 0 & 0\\
  0 & c_{23} & s_{23} \\
  0 & -s_{23} & c_{23}
 \end{pmatrix}
 \begin{pmatrix}
  c_{13} & 0 & s_{13} e^{-i\delta} \\
  0 & 1 & 0\\
  -s_{13} e^{i\delta} & 0 & c_{13}
 \end{pmatrix}
 \begin{pmatrix}
  c_{12} & s_{12} & 0\\
  -s_{12} & c_{12} & 0\\
  0 & 0 & 1
 \end{pmatrix} ,
\end{eqnarray}
where $c_{ij} \equiv \cos\theta_{ij}$
and $s_{ij} \equiv \sin\theta_{ij}$.
 The measurement of the $\nu_\mu^{}$ disappearance
at the T2K experiment~\cite{Abe:2014ugx} shows
\begin{eqnarray}
\sin^2\theta_{23} = 0.514^{+0.055}_{-0.056} , \quad
\Delta m^2_{32} = (2.51 \pm 0.10)\times 10^{-3}\,\eV^2 ,
\end{eqnarray}
for the normal mass ordering~($m_1 < m_3$)
and 
\begin{eqnarray}
\sin^2\theta_{23} = 0.511^{+0.055}_{-0.055} , \quad
\Delta m^2_{23} = (2.48 \pm 0.10)\times 10^{-3}\,\eV^2 ,
\end{eqnarray}
for the inverted mass ordering~($m_3 < m_1$).
 A combined analysis~\cite{Aharmim:2011vm}
of the solar neutrino measurements
and the KamLAND data results in
\begin{eqnarray}
\tan^2\theta_{12} = 0.427^{+0.027}_{-0.024} , \quad
\Delta m^2_{21} = 7.46^{+0.20}_{-0.19}\times 10^{-5}\,\eV^2 .
\end{eqnarray}
 The reactor $\overline{\nu_e^{}}$ measurement
at the DayaBay experiment~\cite{An:2015rpe} gives
\begin{eqnarray}
\sin^2{2\theta_{13}} = 0.084 \pm 0.005 .
\end{eqnarray}

 Let us define $(Y_A)_{e\mu} \equiv (m_\tau/m_\mu)^2 (Y_A^\prime)_{e\mu}$
and keep only terms of $m_\tau^2$ in eq.~\eqref{eq:mnu}.
 We then obtain $(m_\nu^{})_{\tau\tau} = 0$,
which can be satisfied for the inverted mass ordering
with the following values:
\begin{eqnarray}
 &&
  \sin^2\theta_{23} = 0.511 , \quad
  \sin^2{2\theta_{13}} = 0.09 , \quad
  \tan^2\theta_{12} = 0.427 ,
\\
 &&
  \alpha_{12} = \alpha_{32} = \pi , \quad
  \delta = 0 ,
\\
 &&
  \Delta m^2_{23} = 2.48 \times 10^{-3}\,\eV^2 , \quad
  \Delta m^2_{21} = 7.46 \times 10^{-5}\,\eV^2 , \quad
  m_3 = 5.05\times 10^{-2}\,\eV .
\end{eqnarray}
 For these values,
the structure of the neutrino mass matrix is given by
\begin{eqnarray}
 m_\nu
 =
  \begin{pmatrix}
   -2.86 &  4.23 & -4.81\\
    4.23 & -2.14 & -3.91\\
   -4.81 & -3.91 &  0
  \end{pmatrix} \times 10^{-2}\,\eV .
\label{eq:mnu-benchmark}
\end{eqnarray}
%
%
%
\begin{table}[t]
\begin{center}
\begin{tabular}{l||l|l}
{}
 & Current bound
 & Future sensitivity
\\\hline\hline
$\text{BR}(\mu \to e\gamma)$
 & $<5.7\times 10^{-13}$ (MEG)~\cite{Adam:2013mnn}
 & $6\times 10^{-14}$ (MEG upgrade)~\cite{Baldini:2013ke}
\\\hline
$\text{BR}(\mu \to \overline{e}ee)$
 & $< 1.0\times 10^{-12}$ (SINDRUM)~\cite{Bellgardt:1987du}
 & $\sim 10^{-16}$ (Mu3e)~\cite{Blondel:2013ia}
\\\hline\hline
$\text{BR}(\tau \to e\gamma)$
 & $<1.2\times 10^{-7}$ (Belle)~\cite{Hayasaka:2007vc}
 & 
\\
 {}
 & $<3.3\times 10^{-8}$ (Babar)~\cite{Aubert:2009ag}
 &
\\\hline
$\text{BR}(\tau \to \mu\gamma)$
 & $<4.5\times 10^{-8}$ (Belle)~\cite{Hayasaka:2007vc}
 & $5\times 10^{-9}$ (Belle II)~\cite{Abe:2010gxa}
\\
 {}
 & $<4.4\times 10^{-8}$ (Babar)~\cite{Aubert:2009ag}
 & 
\\\hline
$\text{BR}(\tau \to \overline{e}ee)$
 & $< 2.7\times 10^{-8}$ (Belle)~\cite{Hayasaka:2010np}
 & 
\\
 {}
 & $< 2.9\times 10^{-8}$ (Babar)~\cite{Lees:2010ez}
 &
\\\hline
$\text{BR}(\tau \to \overline{e}e\mu)$
 & $< 1.8\times 10^{-8}$ (Belle)~\cite{Hayasaka:2010np}
 &
\\
 {}
 & $< 2.2\times 10^{-8}$ (Babar)~\cite{Lees:2010ez}
 &
\\\hline
$\text{BR}(\tau \to \overline{e}\mu\mu)$
 & $< 1.7\times 10^{-8}$ (Belle)~\cite{Hayasaka:2010np}
 &
\\
 {}
 & $< 2.6\times 10^{-8}$ (Babar)~\cite{Lees:2010ez}
 &
\\\hline
$\text{BR}(\tau \to \overline{\mu}ee)$
 & $< 1.5\times 10^{-8}$ (Belle)~\cite{Hayasaka:2010np}
 &
\\
 {}
 & $< 1.8\times 10^{-8}$ (Babar)~\cite{Lees:2010ez}
 &
\\\hline
$\text{BR}(\tau \to \overline{\mu}e\mu)$
 & $< 2.7\times 10^{-8}$ (Belle)~\cite{Hayasaka:2010np}
 &
\\
 {}
 & $< 3.2\times 10^{-8}$ (Babar)~\cite{Lees:2010ez}
 &
\\\hline
$\text{BR}(\tau \to \overline{\mu}\mu\mu)$
 & $< 2.1\times 10^{-8}$ (Belle)~\cite{Hayasaka:2010np}
 & $1\times 10^{-9}$ (Belle II)~\cite{Abe:2010gxa}
\\
 {}
 & $< 3.3\times 10^{-8}$ (Babar)~\cite{Lees:2010ez}
 &
\\
 {}
 & $< 4.6\times 10^{-8}$ (LHCb)~\cite{Aaij:2014azz}
 & 
\\\hline
$\text{BR}(\tau \to \mu \eta)$
 & $< 6.5\times 10^{-8}$ (Belle)~\cite{Miyazaki:2007jp}
 & 
\end{tabular}
\end{center}
\caption{
 Constraints and future sensitivities
for LFV decays of charged leptons
at 90\,\% confidence level~(CL).
}
\label{tab:LFV}
\end{table}
 When we search for a set of model parameters
which gives the structure of $m_\nu$ in eq.~\eqref{eq:mnu-benchmark},
constraints from LFV decays of charged leptons
have to be taken into account.
 In Table~\ref{tab:LFV},
we summarize current data from various LFV experiments.


 By assuming
$X_{\mu e} = 0$, $X_{\mu\mu} = X_{\tau\tau}$,
and $X_{\mu\tau} = X_{\tau\mu}$,
the neutrino mass matrix in eq.~\eqref{eq:mnu}
is determined by
five combinations of model parameters:
$X_{\tau e}/X_{\tau\tau}$, $X_{\tau\mu}/X_{\tau\tau}$,
$(Y_A)_{e\tau}/(Y_A)_{\mu\tau}$, $(Y_A^\prime)_{e\mu}/(Y_A)_{\mu\tau}$,
and $C_\text{loop} m_\tau^2 (Y_A)_{\mu\tau} X_{\tau\tau}$.
 They are constrained as
\begin{eqnarray}
 \frac{ (Y_A)_{e\tau} }{ (Y_A)_{\mu\tau} }\,
 \frac{ X_{\tau e} }{ X_{\tau\tau} }
 &=&
  \frac{ (m_\nu)_{ee} }{ 2 (m_\nu)_{\mu\tau} }
 =
  0.366 ,
\label{eq:X31}
\\
%
%
 \frac{ (Y_A^\prime)_{e\mu} }{ (Y_A)_{\mu\tau} }\,
 \frac{ X_{\tau\mu} }{ X_{\tau\tau} }
 +
 \frac{ (Y_A)_{e\tau} }{ (Y_A)_{\mu\tau} }
 &=&
  \frac{ (m_\nu)_{e\tau} }{ (m_\nu)_{\mu\tau} }
 =
  1.23 ,
\label{eq:Y13}
\\
%
%
 \frac{ (Y_A^\prime)_{e\mu} }{ (Y_A)_{\mu\tau} }
 +
 \frac{ (Y_A)_{e\tau} }{ (Y_A)_{\mu\tau} }\,
 \frac{ X_{\tau\mu} }{ X_{\tau\tau} }
 +
 \frac{ X_{\tau e} }{ X_{\tau\tau} }
 &=&
  \frac{ (m_\nu)_{e\mu} }{ (m_\nu)_{\mu\tau} }
 =
  -1.08 ,
\label{eq:Yp}
\\
%
%
 \frac{ X_{\tau\mu} }{ X_{\tau\tau} }
 &=&
  \frac{ (m_\nu)_{\mu\mu} }{ 2 (m_\nu)_{\mu\tau} }
 =
  0.273 ,
\label{eq:X32}
\\
%
%
 \frac{\sqrt{2}}{v} C_\text{loop} m_\tau^2 (Y_A)_{\mu\tau} X_{\tau\tau}
 &=&
  (m_\nu^{})_{\mu\tau}
 =
  -3.91 \times 10^{-2}\,\eV .
\label{eq:CYX}
\end{eqnarray}
 These constraints are satisfied
with the following benchmark set for model parameters%
\footnote{
 If the contribution from $(Y_A)_{e\mu}$ to $m_\nu$
is naively ignored,
larger off-diagonal elements of $X$
are necessary.
 We do not take this option
in order to suppress
LFV decays of charged leptons
as much as possible.
}:
\begin{eqnarray}
&&
 M_0 \ \equiv \
 m_{\wt{L}}^{} =  m_{\wt{\ell}}^{}
 = M_1 
 = 10\,\TeV , \quad
 M_2 = 3\,\TeV ,\\
&&
 (\Delta m_{\wt{L}}^2)_{\tau\mu}
 = (\Delta m_{\wt{\ell}}^2)_{\tau\mu}
 = -(0.8\,M_0)^2 ,\\
&&
 (\Delta m_{\wt{L}}^2)_{\tau e}
 = (\Delta m_{\wt{\ell}}^2)_{\tau e}
 = -(0.708\,M_0)^2 ,\\
&&
 (\Delta m_{\wt{L}}^2)_{\mu e}
 = (\Delta m_{\wt{\ell}}^2)_{\mu e}
 = 0 ,
\label{eq:slmass21}\\
&&
 \mu_\Phi^{} = 762\,M_0, \quad
 \tan\beta = 2 ,\\
&&
 m_A = 380\,\GeV , \quad
 m_{H_1^+}^{} = 350\,\GeV , \quad
 \sin^2{\theta_+} = 10^{-5} ,\\
&&
 (Y_A)_{\mu\tau}
  = -1.31\times 10^{-4} , \quad
 (Y_A)_{e\tau}
   = -2.24\times 10^{-4} ,
\label{eq:YA-1}\\
&&
 (Y_A)_{e\mu}
  = \frac{ m_\tau^2 }{ m_\mu^2 } (Y_A^\prime)_{e\mu}
  = 6.50\times 10^{-1} ,
\label{eq:YA-2}
\end{eqnarray}
where
$m_{H_2^+}^2 \simeq m_A^2 + m_W^2 = (388\,\GeV)^2$
at the tree level.
 For the mixing angle $\alpha$ of CP-even neutral Higgs bosons,
we obtain $\cos(\beta-\alpha) = -0.027$
from $\tan(2\alpha)/\tan(2\beta) = (m_A^2 + m_Z^2)/(m_A^2 - m_Z^2)$.
 Notice that
a low $\tan\beta$ value and
large scales of $\mu_\Phi^{}$ and soft breaking parameters
in the benchmark set
are chosen in order to satisfy
constraints from neutrino oscillation data
and searches for the LFV decays of charged leptons.
 Nevertheless,
even taking such a low $\tan\beta$ value,
it can also be compatible with $m_h = 125\,\GeV$
if we take
$M_S^2 \equiv m_{\wt{t}_1}^{} m_{\wt{t}_2}^{} > (O(10)\,\TeV)^2$%
~\cite{Djouadi:2013vqa},
where $ m_{\wt{t}_1}^{}$ and $ m_{\wt{t}_2}^{}$
are masses of two stops,
and such a large value of $M_S$ is consistent with
large breaking scales in the benchmark set.

 The value $m_A = 380\,\GeV$
satisfies the constraint $m_A \gtrsim 350\,\GeV$
for $\tan\beta = 2$~\cite{Djouadi:2015jea}
which comes from the $A^0 \to Z h^0$ search
at the CMS experiment
with the $19.7\,\text{fb}^{-1}$ data
at $\sqrt{s} = 8\,\TeV$~\cite{CMS:2014yra}.
 The values at the benchmark point for $m_A$ and $\tan\beta$
also satisfy the constraint at the ATLAS experiment
with the $20.3\,\text{fb}^{-1}$ data
at $\sqrt{s} = 8\,\TeV$~\cite{Aad:2015wra}.
 The value $m_{H_1^+}^{} = 350\,\GeV$ for $H_1^+ \simeq \omega_1^+$
is consistent with a constraint on $m_{\wt{\ell}}^{}$
for a massless neutralino,
which is $m_{\wt{\ell}}^{} \gtrsim 250\,\GeV$
obtained at the ATLAS experiment
for $20.3\,\text{fb}^{-1}$ of the integrated luminosity
with $\sqrt{s}= 8\,\TeV$~\cite{Aad:2014vma}.
 The CMS experiment gives
$m_{\wt{\ell}}^{} \gtrsim 200\,\GeV$
for $19.5\,\text{fb}^{-1}$ of the integrated luminosity
with $\sqrt{s}= 8\,\TeV$~\cite{Khachatryan:2014qwa}.

 The Yukawa matrix for $\Phi_0^\ast$ in eq.~\eqref{eq:Yukawa-2}
with the benchmark set is calculated as
\begin{eqnarray}
X^\prime
 \equiv
  \frac{\sqrt{2}}{v} M_\ell X
 =
 \begin{pmatrix}
  -1.63 \times 10^{-5}
   & 0
   & -3.49 \times 10^{-6}\\
  0
   & -3.39 \times 10^{-3}
   & 9.25 \times 10^{-4}\\
  -1.22 \times 10^{-2}
   & -1.55 \times 10^{-2}
   & -5.69 \times 10^{-2}
 \end{pmatrix} .
\label{eq:Xbenchmark}
\end{eqnarray}
 This matrix is the source of FCNC\@.

\section{Phenomenology}
\label{sec:pheno}

\subsection{Lepton Flavor Violating Decays of Charged Leptons}

 For $\ell \to \ellp \gamma$,
contributions from one-loop diagrams with charged scalars $H_i^\pm$
are negligible
because $(Y_A)_{\ell \ell^{\prime\prime}} (Y_A)_{\ell^{\prime\prime}\ellp}$
are small enough.
 Since slepton masses are $O(10)\,\TeV$,
one-loop diagrams involving sleptons
have only negligible contributions to $\ell \to \ell^\prime \gamma$
although $(\Delta m_{\wt{L}}^2)_{\tau e}$ and $(\Delta m_{\wt{L}}^2)_{\tau\mu}$
have a similar size to $m_{\wt{L}}^2$~(also for $m_{\wt{\ell}}^2$).
 The condition in eq.~\eqref{eq:slmass21}
forbids not only the one-loop contribution of sleptons
to $\mu \to e\gamma$
but also the Barr-Zee type two-loop contributions%
~\cite{Barr:1990vd,Hisano:2010es}
with $X_{\mu e}^\prime$.
 The dominant contribution to $\mu \to e\gamma$
comes from a one-loop diagram involving $\tau$
with $X_{\tau\mu}^\prime X_{\tau e}^\prime$,
which results in $\BR(\mu \to e\gamma) = 1.5\times 10^{-13}$
with $m_H \simeq m_A$.
 This value satisfies the current constraint
$\BR(\mu \to e\gamma) < 5.7\times 10^{-13}$~(90\,\%~CL)
at the MEG experiment~\cite{Adam:2013mnn},
and it could be observed
at a planned MEG experiment upgrade~\cite{Baldini:2013ke}
where a sensitivity for $\BR(\mu \to e\gamma) \simeq 6\times 10^{-14}$
is expected.
 If we take a larger value of $\tan\beta$~(e.g., $\tan\beta = 3$),
$X_{\tau\tau}^\prime$ is enhanced,
and then it is required that
$X_{\tau\mu}^\prime$~(and also $X_{\tau e}^\prime$ in fact)
becomes larger due to the condition in eq.~\eqref{eq:X32};
 thus, a low $\tan\beta$ value is required
to satisfy the constraint on $\BR(\mu \to e\gamma)$
with $m_A = 380\,\GeV$ which is experimentally accessible.
 On the other hand,
Barr-Zee diagrams involving the top quark in a loop
give dominant contributions
to $\tau \to \ell \gamma$ because
$(\Delta m^2_{\wt{L}})_{\tau e}$,
$(\Delta m^2_{\wt{\ell}})_{\tau e}$,
$(\Delta m^2_{\wt{L}})_{\tau\mu}$,
and $(\Delta m^2_{\wt{\ell}})_{\tau\mu}$
are not zero at our benchmark set.
 By using formulae in e.g.\ Ref.~\cite{Hisano:2010es}
with $\BR(\tau \to e\nu_\tau\overline{\nu}_e) = 0.17$,
we have $\BR(\tau \to e\gamma) = 2.7\times 10^{-9}$
and $\BR(\tau \to \mu\gamma) = 4.4\times 10^{-9}$
which satisfy
$\BR(\tau \to e\gamma) < 3.3\times 10^{-8}$~(90\,\%~CL)
and $\BR(\tau \to \mu\gamma) < 4.4\times 10^{-8}$~(90\,\%~CL)
obtained at the Babar experiment~\cite{Aubert:2009ag}.
 These values are comparable to
expected sensitivity $\BR(\tau \to \ell \gamma) \sim 10^{-9}$
at the Belle~II experiment~\cite{Abe:2010gxa}.

 The Yukawa matrix $X^\prime$ can cause $\mu \to \overline{e}ee$
and $\tau \to \overline{\ell}\ellp\ell^{\prime\prime}$
at the tree level.
 Although
there is the stringent experimental constraint
$\BR(\mu \to \overline{e}ee) < 1.0\times 10^{-12}$~(90\,\%~CL)
at the SINDRUM experiment~\cite{Bellgardt:1987du},
our benchmark set trivially satisfies it
because $X_{\mu e}^\prime = 0$ in eq.~\eqref{eq:Xbenchmark}
gives $\BR(\mu \to \overline{e}ee) = 0$ at the tree level.
 For $\tau \to \overline{\ell}\ellp\ell^{\prime\prime}$,
the tree level contributions at the benchmark set result in
$\BR(\tau \to \overline{e}\mu\mu)
= \BR(\tau \to \overline{\mu}ee) = 0$,
$\BR(\tau \to \overline{e}ee)
\sim \BR(\tau \to \overline{e}e\mu) \sim 10^{-16}$,
and
$\BR(\tau \to \overline{\mu}e\mu)
\sim \BR(\tau \to \overline{\mu}\mu\mu) \sim 10^{-11}$
where
experimental bounds are
$\BR(\tau \to \overline{\ell}\ellp\ell^{\prime\prime}) \lesssim 10^{-8}$%
~(90\,\%~CL)~\cite{Hayasaka:2010np,Lees:2010ez,Aaij:2014azz}.
 The constraint
$\BR(\tau \to \mu \eta) < 6.5 \times 10^{-8}$~(90\,\%~CL)%
~\cite{Miyazaki:2007jp}
is also satisfied
even if we use the relation
$\BR(\tau \to \mu \eta)
\simeq 8.4\times \BR(\tau \to \overline{\mu}\mu\mu)$%
~\cite{Sher:2002ew}.

 The LFV coupling $X^\prime_{\tau\mu} = -1.55 \times 10^{-2}$
in Eq.~\eqref{eq:Xbenchmark}
is comparable to $\sqrt{2} m_\tau/v = 10^{-2}$.
 However,
the branching ratio $\BR(h \to \mu\tau)$
is suppressed to about $10^{-4}$
by $\cos^2(\alpha - \beta) \simeq 10^{-3}$.
 Therefore,
the benchmark point satisfies
the current upper limit on $\BR(h \to \mu\tau)$%
~\cite{Khachatryan:2015kon, Aad:2015gha},
although the $2.4\,\sigma$ excess
which is currently reported by the CMS~\cite{Khachatryan:2015kon}
is not explained by our benchmark point.

\subsection{Dark Matter}

 Large values of $\mu_\Phi^{}$ and $M_1$ are preferred
in order to obtain sizable
off-diagonal elements of $X$~(namely, $\epsilon_2$)
which are required for the appropriate structure
of the neutrino mass matrix.
 On the other hand,
the value of $M_2$ is not required to be very large.
 Therefore,
there is a possibility of the wino dark matter
in the SUSY Zee model.
 In our benchmark set,
we take $M_2 = 3\,\TeV$
for which the relic abundance of dark matter
can be explained~\cite{Hisano:2006nn}.
 The spin-independent cross section
of the wino scattering on a proton
is evaluated as $\sim 10^{-47}\,\text{cm}^2$%
~(See e.g., Ref.~\cite{Hisano:2015rsa}),
which is greater than the neutrino background~\cite{Billard:2013qya}.


\subsection{Phenomenology of Charged Scalar Bosons}

\begin{figure}[t]
\begin{center}
\includegraphics[scale=0.82]{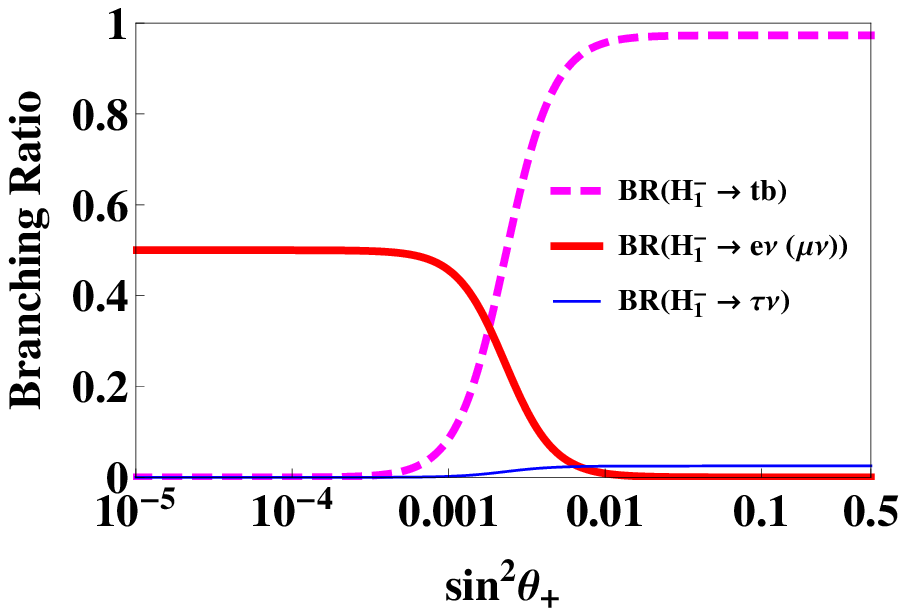} \hspace*{7mm}
\includegraphics[scale=0.82]{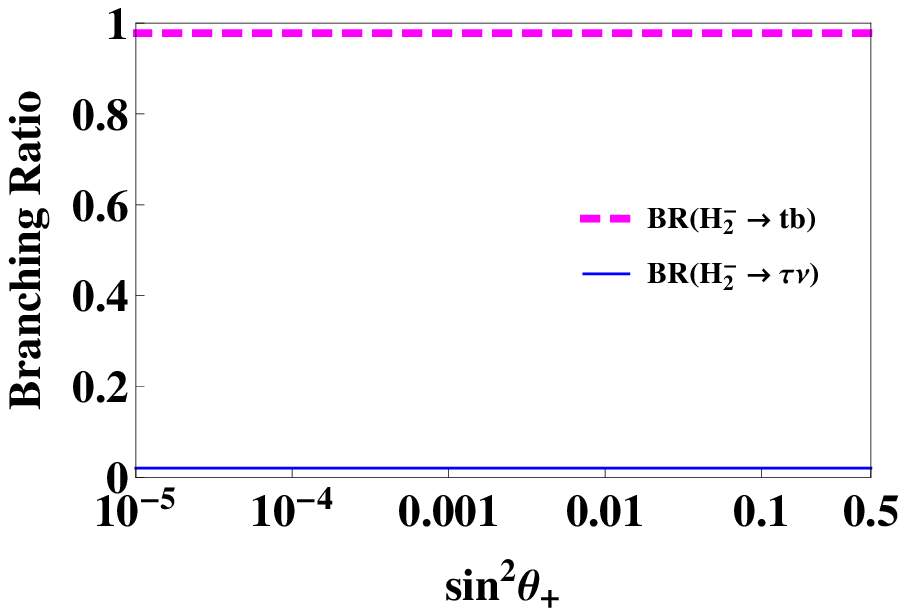}
\vspace*{-5mm}
\caption{
 Decay branching ratios of $H_1^\pm$~(left) and $H_2^\pm$~(right)
with respect to $\sin^2\theta_+$.
 The red thick and the blue thin solid lines show
$\BR(H^\pm \to e\nu)$~($= \BR(H^\pm \to \mu\nu)$)
and $\BR(H^- \to \tau\nu)$, respectively.
 The magenta dashed line is for $\BR(H^\pm \to tb)$.
}
\label{fig:BR}
\end{center}
\end{figure}

 Decay branching ratios of $H_1^\pm$ and $H_2^\pm$
are shown in Fig.~\ref{fig:BR} as a function of $\sin^2\theta_+$
where $H_1^\pm = \omega_1^\pm$ if $\sin^2\theta_+ = 0$.
 The red thick and the blue thin solid lines show
$\BR(H^\pm \to e\nu)$~($= \BR(H^\pm \to \mu\nu)$)
and $\BR(H^\pm \to \tau\nu)$, respectively.
 The magenta dashed line is for $\BR(H^\pm \to tb)$.
 Except for $\sin^2\theta_+$ and $Y_A$,
parameters are set to the benchmark point.
 Values of elements of $Y_A$ depend on $\sin^2\theta_+$
through a condition eq.~\eqref{eq:CYX}.
 For $\sin^2\theta_+ \gtrsim 10^{-2}$,
$H_1^-$ is the doublet-like Higgs boson
for which the decay $H_1^- \to tb$
is the dominant channel.
 In this case,
since the dominant decay channel of $H_2^-$
is also $tb$~(for any $\sin^2\theta_+$),
existence of an $\SU(2)$-singlet component is hidden.
 Such doublet-like charged Higgs bosons%
~(the same as the one in the Type-II two Higgs doublet model)
with $m_{H_1^\pm}^{} = 350\,\GeV$ and $m_{H_2^\pm}^{} \simeq 388\,\GeV$
for $\tan\beta = 2$
may be observed at the LHC with $\sqrt{s} = 14\,\TeV$
and $300\,\text{fb}^{-1}$ of the integrated luminosity
via the production process $g b \to t H^\pm$
followed by the decay $H^\pm \to tb$%
~\cite{Borzumati:1999th, Kanemura:2014dea}.

 On the other hand,
$H_1^\pm$ dominantly decays into leptons via $Y_A$
for $\sin^2\theta_+ \lesssim 10^{-3}$.
 The hierarchical structure of $Y_A$
in eqs.~\eqref{eq:YA-1} and \eqref{eq:YA-2}
gives a characteristic prediction
\begin{eqnarray}
 \BR( H_1^\pm \to e\nu )
  : \BR( H_1^\pm \to \mu\nu ) 
  : \BR( H_1^\pm \to \tau\nu )
 \simeq
  1 : 1 : 0 ,
\label{eq:BR-HP1}
\end{eqnarray}
where neutrinos in the final states are summed
because experiments are not sensitive to their flavors.
 If the coincidence
$\BR( H_1^\pm \to e\nu )  = \BR( H_1^\pm \to \mu\nu )$
is observed,
it would be regarded as a nonaccidental one
but a natural consequence of the $\SU(2)_L$-singlet charged scalar
with a $(Y_A)_{e\mu}$-dominated Yukawa matrix.
 Nonobservation of the signal of $H_1^\pm \to \tau\nu$
would also suggest that
$H_1^\pm$ does not come from an $\SU(2)_L$-doublet Higgs
which has a vacuum expectation value.
 For the singlet-like charged scalar,
a region of $m_{H_1^\pm}^{} \lesssim 430\,\GeV$
can be probed at the LHC with $\sqrt{s} = 14\,\TeV$
and $100\,\text{fb}^{-1}$ of the integrated luminosity~\cite{Eckel:2014dza}.
 At the ILC with $\sqrt{s} = 1\,\TeV$,
the cross section is about $10\,\text{fb}$~\cite{Kanemura:2000bq}
which would be enough to observe $H_1^\pm$.
 The singlet-like $H_1^\pm$
would be distinguished from right-handed sleptons
if the lightest neutralino is sufficiently heavy
such that it does not mimic a neutrino.

\begin{figure}[t]
\begin{center}
\includegraphics[scale=0.82]{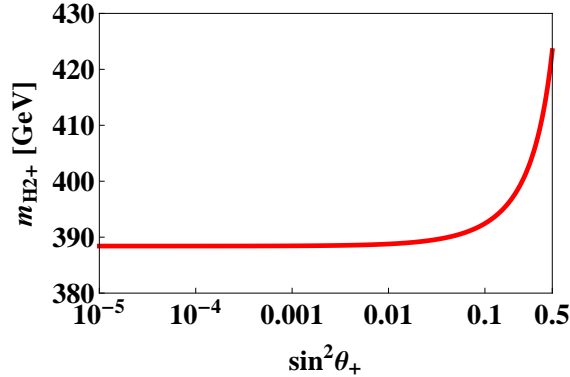}
\vspace*{-5mm}
\caption{
 The mass of $H_2^\pm$ at the tree level
with respect to $\sin^2\theta_+$
for $m_A^{} = 380\,\GeV$ and $m_{H_1^\pm}^{} = 350\,\GeV$.
}
\label{fig:mHP2}
\end{center}
\end{figure}

 The behavior of $m_{H_2^+}^2$ at the tree level
with respect to $\sin^2\theta_+$
is shown in Fig.~\ref{fig:mHP2}.
 In the region of small $\sin^2\theta_+$,
the relation $m_{H_2^+}^2 = m_A^2 + m_W^2$ approximately holds
with an appropriate radiative correction~\cite{Carena:1995bx}.
 On the other hand,
the relation between $m_A$ and
the mass of the charged Higgs boson in the MSSM
at the one-loop level
is expressed as~\cite{Kanemura:2001hz}
\begin{eqnarray}
 m_{H^+}^2
 =
  m_A^2 + m_W^2
  + \Pi_{AA}(m_A^2) - \Pi_{H^+H^-}(m_A^2+m_W^2) + \Pi_{WW}(m_W^2),
\end{eqnarray}
where formulae of self-energies
$\Pi_{AA}(q^2)$, $\Pi_{H^+H^-}(q^2)$, and $\Pi_{WW}(q^2)$
can be found in Ref.~\cite{Kanemura:2001hz}.
 This relation at the one-loop level
also approximately holds for $m_{H_2^+}^2$ in the SUSY Zee model
for a small value of $\sin\theta_+$,
where $H_2^+$ is almost the same as
the charged Higgs boson in the MSSM\@.
 The benchmark set gives $\delta_{H_2^\pm}^{} \simeq 0.1$,
where the $\delta_{H_2^\pm}^{}$ is defined as
$m_{H_2^+}^{} = \sqrt{ m_A^2 + m_W^2 } ( 1 + \delta_{H_2^\pm}^{} )$.
 If the mass relation is experimentally confirmed
in addition to eq.~\eqref{eq:BR-HP1}
which is a consequence of $m_\nu$ in eq.~\eqref{eq:mnu},
the SUSY Zee model can be highly supported.

 The detection prospect of $A^0$ with $m_A = 380\,\GeV$
seems marginal for $A^0 \to \tau\tau$
at the LHC with $\sqrt{s}=14\,\TeV$%
~\cite{Kanemura:2014bqa, Kanemura:2014dea, Djouadi:2015jea}
but sufficient for $A^0 \to Zh$~\cite{Djouadi:2015jea}.
 Discovery of the $A^0$ can be expected
also via $A^0 \to tt$~\cite{Djouadi:2015jea}.
 The cross section for a process
$e^+ e^- \to Z^\ast \to H^0 A^0 \to ttbb$
for the $A^0$ at the ILC with $\sqrt{s} = 1\,\TeV$
is greater than $0.1\,\text{fb}$~\cite{Kanemura:2014dea}
which would be sufficient to detect the signal.

\section{Conclusion}
\label{sec:concl}

 We have extended the Zee model
to a SUSY model with the conserved R-parity.
 The MSSM has been extended by introducing
$\SU(2)_L$-singlet complex superfields
$\wh{\omega}_1^+$ and $\wh{\omega}_2^-$.
 In order to generate three nonzero neutrino masses,
the extension gives the simplest alternative
to the SUSY seesaw model
where three right-handed neutrino superfields are introduced.
 Tiny neutrino masses
are obtained at the one-loop level.
 We have shown that
the mass matrix can be consistent 
with the current neutrino oscillation data
thanks to the nonholomorphic Yukawa interaction,
which is dominantly generated
by one-loop diagrams involving sleptons and the bino.
 We have obtained a benchmark point
which satisfies not only neutrino oscillation data
but also constraints
from LFV decays of charged leptons and
the current LHC results.
 The parameter set is also consistent with $m_h = 125\,\GeV$.
 The dark matter is stabilized
thanks to the R-parity conservation in the SUSY Zee model,
and we have found that the wino can be dark matter.

 While SUSY particles are rather heavy
at the benchmark point,
additional scalar bosons~($H^0$, $A^0$, $H_1^\pm$, and $H_2^\pm$)
are light enough to be discovered
at the LHC and future lepton colliders.
 If the lightest charged scalar bosons $H_1^\pm$
are almost singlet-like due to a small mixing,
their decay branching ratios have been predicted as
$\BR(H_1^\pm \to e\nu) : \BR(H_1^\pm \to \mu\nu)
 : \BR(H_1^\pm \to \tau\nu) = 1 : 1 : 0$.
 On the other hand,
the heavier charged scalar bosons $H_2^\pm$ decay into $tb$.
 For such a small mixing case,
a relation between $m_{H_2^\pm}^{}$ and $m_A$ in the MSSM remains
because $m_{H_2^\pm}^{}$ is almost the same as
the charged Higgs boson mass in the MSSM\@.
 Therefore,
our model can be tested
by measuring the specific decay patterns of $H_1^\pm$ and $H_2^\pm$
and the discriminative mass spectrum of additional scalar bosons.

\begin{acknowledgments}
 We thank Natsumi Nagata for bringing our attention
to the Barr-Zee diagram for $\mu \to e\gamma$.
 This work was supported,
in part, by Grant-in-Aid for Scientific Research,
the Ministry of Education, Culture, Sports, Science and Technology~(MEXT),
No.~24340046 (S.~K.),
and Grant H2020-MSCA-RISE-2014 No.~645722 (Non Minimal Higgs) (S.~K.).
 This work was also supported, in part,
by a Grant-in-Aid for Scientific Research
from the MEXT, Japan, No.~23104011~(T.~S.).
\end{acknowledgments}

\end{document}